\documentclass[vecphys,openany]{svmult}

\usepackage{url}
\usepackage{bm}
\usepackage{subfigure}
\usepackage{makeidx}     

\usepackage{multicol}    
\usepackage[bottom]{footmisc}
\usepackage{latexsym,amsfonts,amsmath,amssymb,graphicx}

\usepackage{epsfig}
\usepackage{colortbl}  
\definecolor{myGray}{gray}{0.9}
\usepackage{fancyvrb}

\begin{document}

\title*{PCA and $K$-Means Decipher Genome}

\author{Alexander N. Gorban\inst{1,3}\and
Andrei Y. Zinovyev\inst{2,3}}
\authorrunning{A.N. Gorban \and
A.Y. Zinovyev}
\institute{University of Leicester, University Road, Leicester, LE1
7RH,  UK, \\ \texttt{ag153@le.ac.uk} \and Institut Curie, 26, rue
d'Ulm, Paris, 75248, France, \\ \texttt{andrei.zinovyev@curie.fr}
\and Institute Of Computational Modeling of Siberian Branch of
Russian Academy of Science, Krasnoyarsk, Russia}
%
%
\maketitle

\begin{abstract}
In this paper, we aim to give a tutorial for undergraduate
students studying statistical methods and/or bioinformatics. The
students will learn how data visualization can help in genomic
sequence analysis. Students start with a fragment of genetic text
of a bacterial genome and analyze its structure. By means of
principal component analysis they ``discover'' that the
information in the genome is encoded by non-overlapping triplets.
Next, they learn how to find gene positions. This  exercise on PCA
and K-Means clustering enables active study of the basic
bioinformatics notions. Appendix 1 contains program listings that
go along with this exercise. Appendix 2 includes 2D PCA plots of
triplet usage in moving frame for a series of bacterial genomes
from GC-poor to GC-rich ones. Animated 3D PCA plots are attached
as separate gif files. Topology (cluster structure) and geometry
(mutual positions of clusters) of these plots depends clearly on
GC-content.
\end{abstract}

\keywords{Bioinfomatics; Data visualization; Cryptography;
Clustering; Principal component analysis}

\section{Introduction}

When it is claimed in newspapers that a new genome is deciphered,
it usually means that the sequence of the genome has been read
only, so that  a long sequence using four genetic letters: A, C, G
and T is known. The first step in complete deciphering of the
genome is identification of the positions of elementary messages
in this text or {\it detecting genes} in the biological language.
This is imperative before trying to understand what these messages
mean for a living cell.

Bioinformatics -- and genomic sequence analysis, in particular --
is one of the hottest topics in modern science. The usefulness of
statistical techniques in this field cannot be underestimated. In
particular, a successful approach to the identification of gene
positions is achieved by statistical analysis of genetic text
composition.

In this exercise, we use Matlab to convert genetic text into a
table of short word frequencies and to visualize this table using
principal component analysis (PCA). Students can find, by
themselves, that the sequence of letters is not random and that
the information in the text is encoded by non-overlapping
triplets. Using the simplest K-Means clustering method from the
Matlab Statistical toolbox, it is possible to detect positions of
genes in the genome and even to predict their direction.

\section{Required Materials}

To follow this exercise, it is necessary to prepare a genomic
sequence. We provide a fragment of the genomic sequence of {\it
Caulobacter Crescentus}. Other sequences can be downloaded from the
\index{Genbank} Genbank FTP-site \cite{Genbank}. Our procedures work
with the files in the {\it Fasta} format (the corresponding files
have a {\it .fa} extension) and are limited to analyzing fragments
of 400--500 kb in length.

Five simple functions are used: \vspace{10pt}

\begin{tabular}{ll}

{\it LoadFreq.m}& loads Fasta-file into a Matlab string\\

{\it CalcFreq.m}& converts a text into a numerical table of short
word \\ &frequencies \\

{\it PCAFreq.m} & visualizes a numerical table using principal
component \\ & analysis \\

{\it ClustFreq.m} & is used to perform clustering with the K-Means
algorithm \\

{\it GenBrowser.m} & is used to visualize the results of
clustering on the text

\end{tabular}
\vspace{10pt}

All sequence files and the m-files should be placed into the current
Matlab working folder. The sequence of \index{Matlab} Matlab
commands for this exercise is the following:

\vspace{15pt}

 \begin{minipage}[l]{10cm}
  {\ttfamily
str = LoadSeq('ccrescentus.fa');

xx1 = CalcFreq(str,1,300);

xx2 = CalcFreq(str,2,300);

xx3 = CalcFreq(str,3,300);

xx4 = CalcFreq(str,4,300);

PCAFreq(xx1);

PCAFreq(xx2);

PCAFreq(xx3);

PCAFreq(xx4);

fragn = ClustFreq(xx3,7);

GenBrowser(str,300,fragn,13000);

 }
 \end{minipage}

\vspace{15pt}

All the required materials can be downloaded from \cite{reqmat}.

\section{Genomic Sequence}

\subsection{Background}

\index{genome} The information that is needed for a living cell
functioning is encoded in a long molecule of DNA. It can be
presented as a text with an alphabet that has only four letters A,
C, G and T. The diversity of living organisms and their complex
properties is hidden in their genomic sequences. One of the most
exciting problems in modern science is to understand the
organization of living matter by reading genomic sequences.

\begin{figure}
\centering{
 \includegraphics[width=115mm, height=60mm]{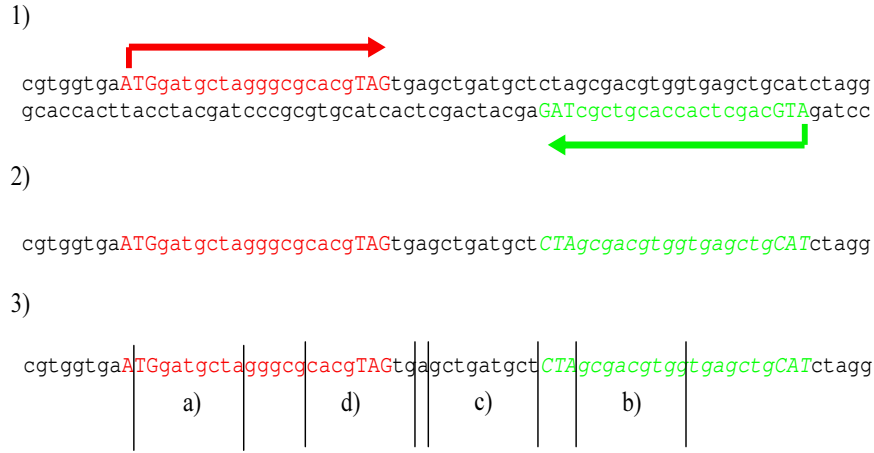}
}\caption{\textbf{1}) DNA can be represented as two complementary
text strings. ``Complementary" here means that instead of any
letter ``a",``t",``g" and ``c" in one string there stands
``t",``a",``c" and ``g",  respectively, in the other string.
Elementary messages or genes can be located in both strings, but
in the lower one they are read from right to the left. Genes
usually start with ``atg" and end with ``tag" or ``taa" or ``tga"
words. \textbf{2}) In databases one obtains only the upper
``forward" string. Genes from the lower ``backward" string can be
reflected on it, but should be read in the opposite direction and
changing the letters, accordingly to the complementary rules.
\textbf{3}) If we take a randomly chosen fragment, it can be of
one of three types: \textbf{a}) entirely in a gene from the
forward string; \textbf{b}) entirely in a gene from the backward
string; \textbf{c}) entirely outside genes; \textbf{d}) partially
in genes and partially outside genes \label{genes}}
\end{figure}

One distinctive message in a genomic sequence is a piece of text,
called {\it a gene}. Genes can be oriented in the sequence in the
forward and backward directions (see Fig.~\ref{genes}). This
simplified picture with unbroken genes is close to reality for
bacteria. In the highest organisms  (humans, for example), the
notion of a gene is more complex.

It was one of many great discoveries of the twentieth century that
biological information is encoded in genes by means of triplets of
letters, called {\it codons} in the biological literature. In the
famous paper by Crick {\it et al}. \cite{14Crick61}, this fact was
proven by genetic experiments carried out on bacteria mutants. In
this exercise, we will analyze this by using the genetic sequence
only.

In nature, there is a special mechanism that is designed to read
genes. It is evident that as the information is encoded by
non-overlapping triplets, it is important for this mechanism to
start reading a gene without a shift, from the first letter of the
first codon to the last one; otherwise, the information decoded
will be completely corrupted.

An easy introduction to modern molecular biology can be found in
\cite{MBsimple}.

\subsection{Sequences for the Analysis}

The work starts with a fragment of genomic sequence of the {\it
Caulobacter Crescentus} bacterium. A short biological description of
this bacterium can be found in \cite{caulodesc}. The sequence is
given as a long text file (300 kb), and the students are asked to
look at the file and ensure that the text uses the alphabet of four
letters ($a$, $c$, $g$ and $t$) and that these letters are used
without spaces. It is noticeable that, although the text seems to be
random, it is well organized, but we cannot understand it without
special tools. Statistical methods can help us understand the text
organization.

The sequence can be loaded in the Matlab environment by {\it
LoadSeq} function:

\vspace{15pt}

\begin{minipage}[l]{10cm}
  {\ttfamily
str = LoadSeq('ccrescentus.fa'); }
\end{minipage}

\section{Converting Text to a Numerical Table}

A {\it word} is any continuous piece of text that contains several
subsequent letters. As there are no spaces in the text, separation
into words is not unique.

The method we use is as follows. We clip the whole text into
fragments of 300 letters\footnote{Mean gene size in bacteria is
about 1000 genetic letters, the fragment length in 300 letters
corresponds well to detect genes on this scale, with some
resolution} in length and calculate the frequencies of short words
(of length 1--4) inside every fragment. This will give us a
description of the text in the form of a numerical table. There will
be four such tables for every short word length from 1 to 4.

As there are only four letters, there are four possible words of
length $1$ (singlets), $16=4^2$ possible words of length $2$
(duplets), $64=4^3$ possible words of length $3$ (triplets) and
$256=4^4$ possible words of length $4$ (quadruplets). The first
table contains four columns (frequency of every singlet) and the
number of rows equals the number of fragments. The second table
has 16 columns and the same number of rows, and so on.

To calculate the tables, students use the {\it CalcFreq.m}
function. The first input argument for the function {\it CalcFreq}
is the string containing the text, the second input argument is
the length of the words to be counted, and the third argument is
the fragment length. The output argument is the resulting table of
frequencies. Students use the following set of commands to
generate tables corresponding to four different word lengths:

\vspace{15pt}

\begin{minipage}[l]{10cm}
  {\ttfamily
xx1 = CalcFreq(str,1,300);

xx2 = CalcFreq(str,2,300);

xx3 = CalcFreq(str,3,300);

xx4 = CalcFreq(str,4,300); }
\end{minipage}

\section{Data Visualization}

\subsection{Visualization}

{\it PCAFreq.m} function has only one input argument, the table
obtained from the previous step. It produces a PCA plot for this
table (PCA plot shows distribution of points on a principal plane,
with the $x$ axis corresponding to the projection of the point on
the first principal component and the $y$ axis corresponding to
the projection on the second one). For an introduction to PCA, see
\cite{PCAuser}.

\begin{figure}
\centering{
\begin{tabular}{lr}

\textbf{a}) \includegraphics[width=45mm, height=45mm]{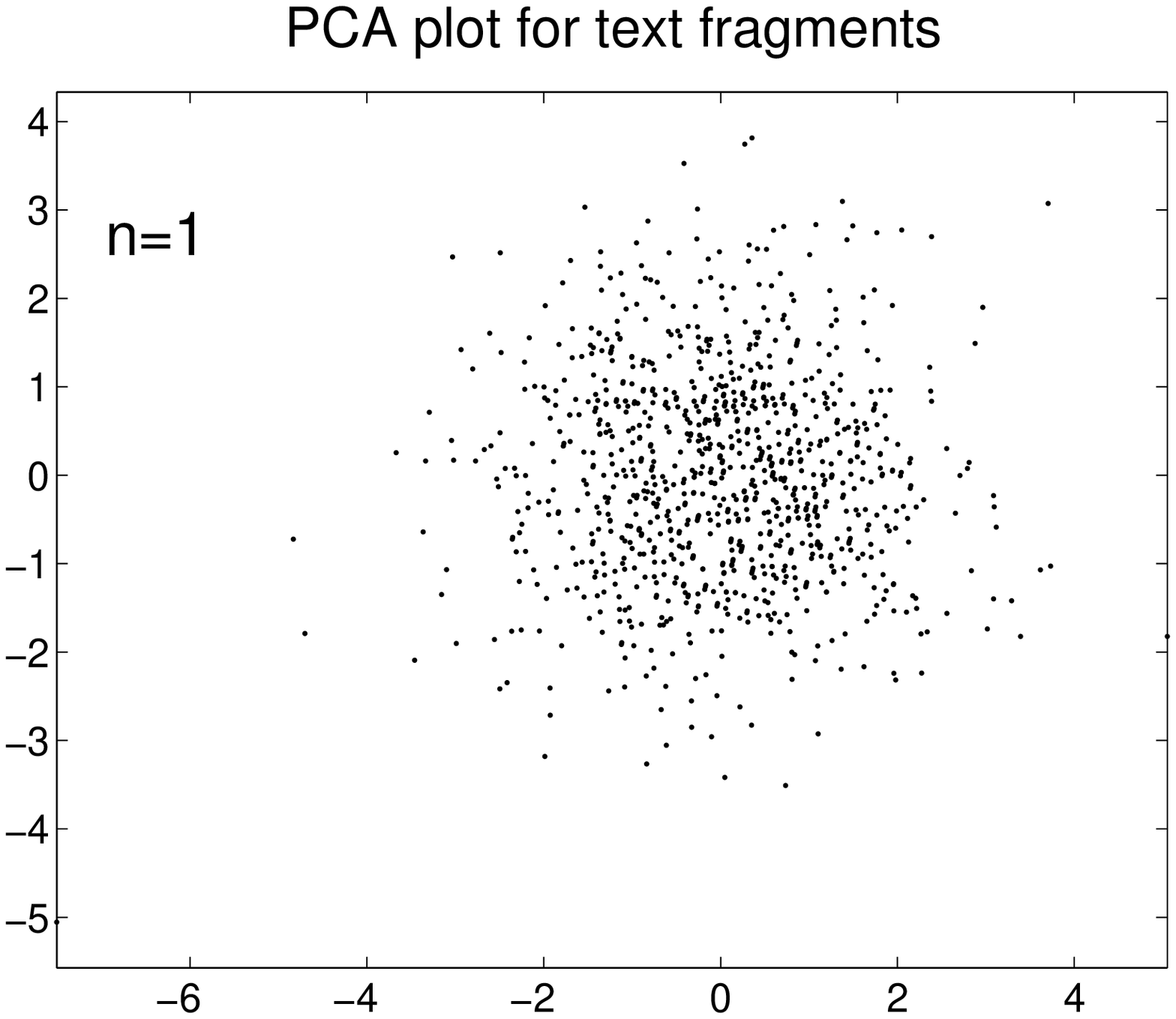}
 &
\textbf{b}) \includegraphics[width=45mm, height=45mm]{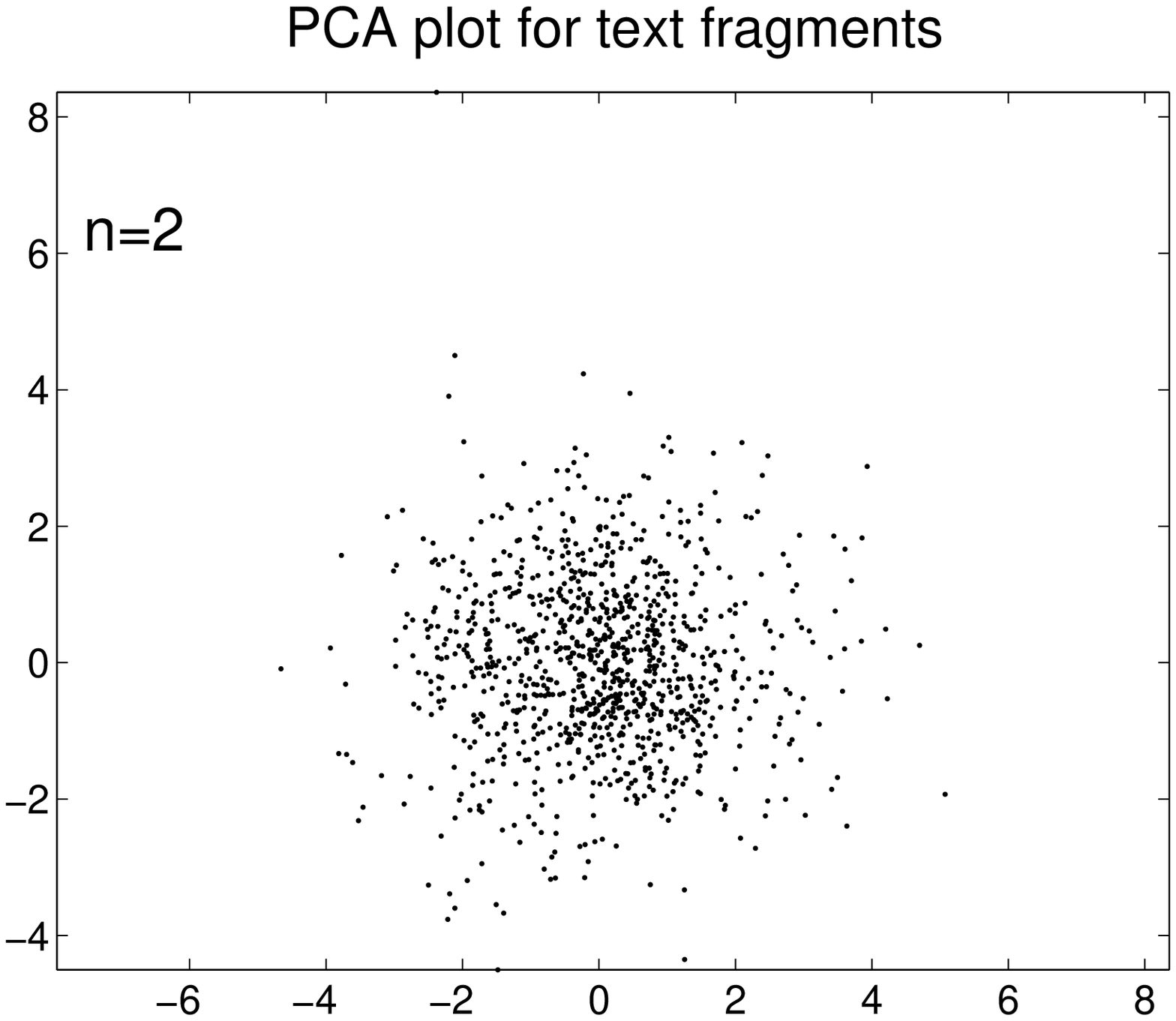}
  \\
\textbf{c}) \includegraphics[width=45mm, height=45mm]{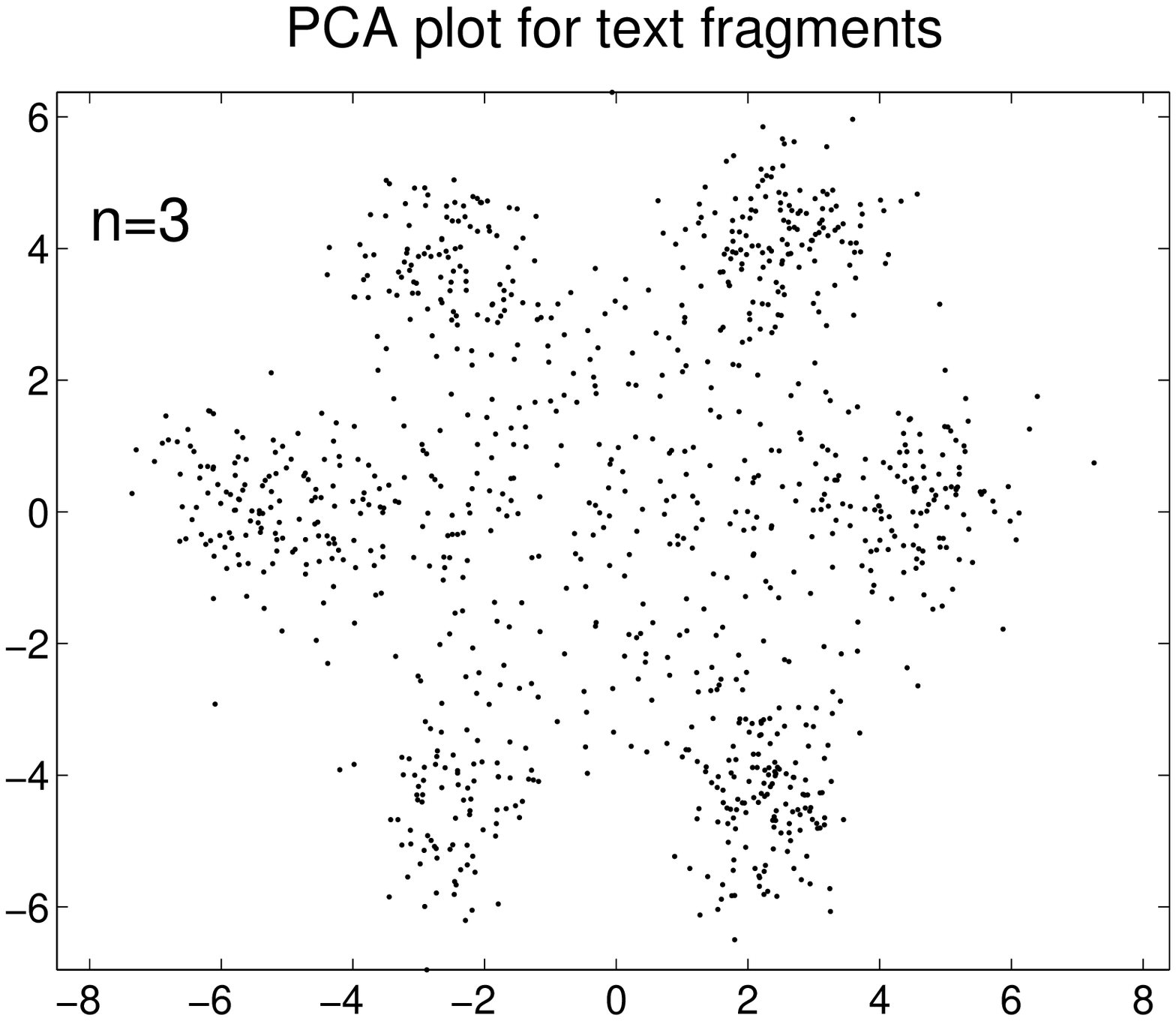}
 &
\textbf{d}) \includegraphics[width=45mm, height=45mm]{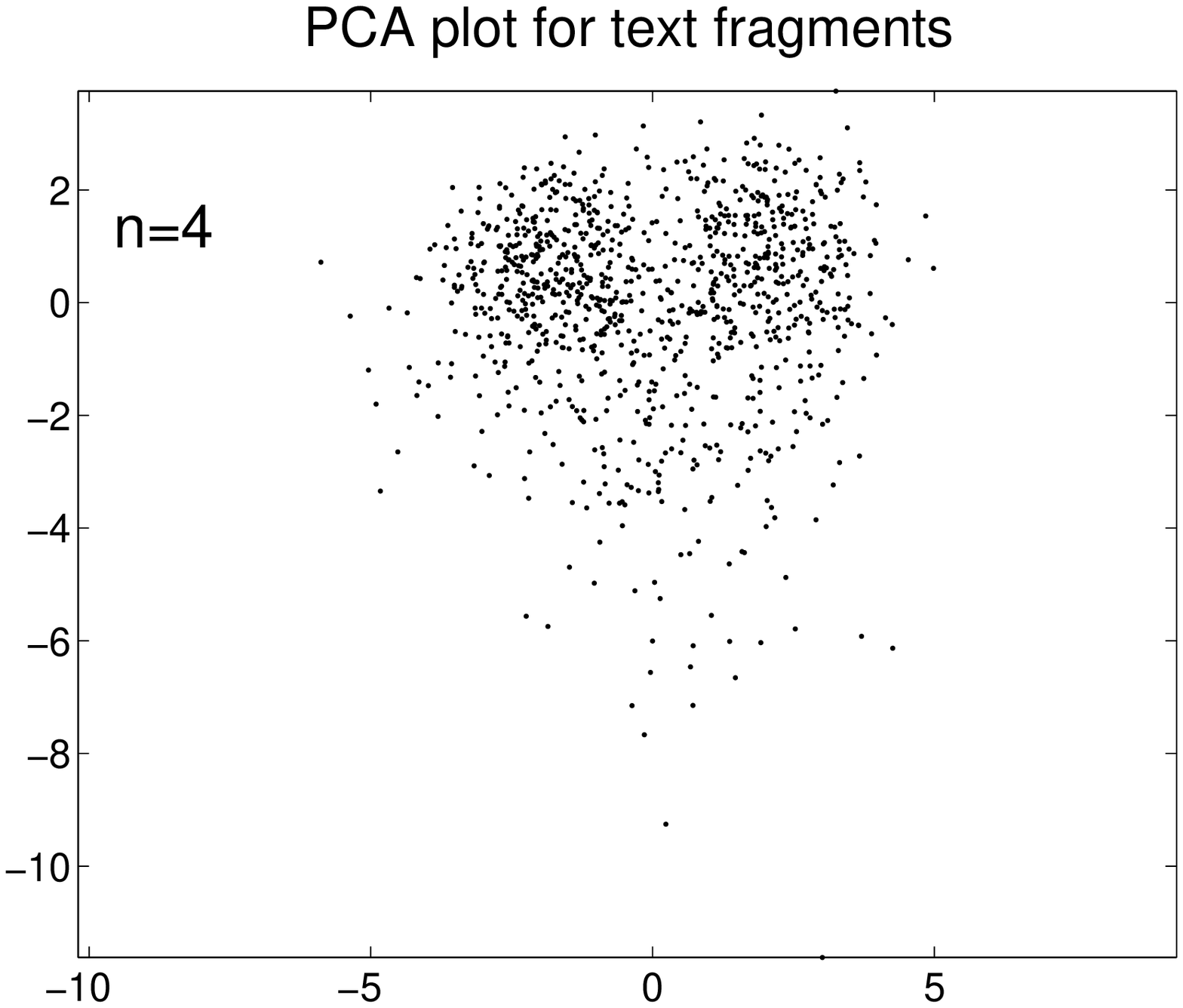}

\end{tabular}
}\caption{PCA plots of word frequencies of different length.
\textbf{c})~Shows the most structured distribution. The structure
can be interpreted as the existence of a non-overlapping triplet
code \label{14pcas}}
\end{figure}

By typing the commands below, students produce four plots for the
word lengths from 1 to 4 (see Fig.~\ref{14pcas}).

\vspace{15pt}

\begin{minipage}[l]{10cm}
  {\ttfamily
PCAFreq(xx1);

PCAFreq(xx2);

PCAFreq(xx3);

PCAFreq(xx4); }
\end{minipage}

\subsection{Understanding Plots}

The main message in these four pictures is that the genomic text
contains information that is encoded by non-overlapping {\it
triplets}, because the plot corresponding to the triplets is
evidently highly structured as opposed to the pictures of
singlets, duplets and quadruplets. The triplet picture evidently
contains 7 clusters.

It is important to explain to students how 7-cluster structure in
Fig.~\ref{genes} occurs in nature.

Let us suppose that the text contains genes and that information is
encoded by non-overlapping subsequent triplets (codons), but we do
not know where the genes start and end or what the frequency
distribution of codons is.

Let us blindly cut the text into fragments. Any fragment can
contain: a) a piece of a gene in the forward direction; b) a piece
of a gene in the backward direction; c) no genes (non-coding
part); d) or a mix of coding and non-coding parts.

Consider the first case (a). The fragment can overlap with a gene
in three possible ways, with three possible shifts (mod3) of the
first letter of the fragment with respect to the start of the
gene. If we enumerate the letters in the gene form the first one,
1-2-3-4-..., then the first letter of the fragment can be in the
sequence 1-4-7-10-... (=1(mod(3)) (a ``correct" shift), the
sequence 2-5-8-11-... (=2(mod(3)), or 3-6-9-12-... (=0(mod(3)). If
we start to read the information triplet by triplet, starting from
the first letter of the fragment, we can read the gene correctly
only if the fragment overlaps with it with a correct shift (see
Fig.~\ref{genes}). In general, if the start of the fragment is not
chosen deliberately, then we can read the gene in three possible
ways. Therefore, this first case (a) generates three possible
frequency distributions, each one of which is ``shifted" in
relation to another.

The second case (b) is analogous and also gives three possible
triplet distributions. They are not independent of the ones
obtained in the first cases for the following reason. The
difference is the triplets are read ``from the end to the
beginning'' which produces a kind of mirror reflection of the
triplet distributions from the first case (a).

The third case (c) produces only one distribution, which is
symmetrical with respect to the `shifts' (or rotations) in the
first two cases, and there is a hypothesis that this is a result
of genomic sequence evolution. This can be explained as follows.

Vitality of a bacterium depends on the correct functioning of all
biological mechanisms. These mechanisms are encoded in genes, and if
something wrong happens with gene sequences (for example there is an
error when DNA is duplicated), then the organism risks becoming
non-vital. Nothing is perfect in our world and the errors happen all
the time, including during DNA duplication. These \index{mutations}
errors are called {\it mutations}.

The most dangerous mutations are those that change the reading
frame, i.e. letter deletions or insertions. If such a mutation
happens inside a gene sequence, the rest of the gene becomes
corrupted: the reading mechanism (which reads the triplets one by
one and does not know about the mutation) will read it with a shift.
Because of this the organisms with such mutations often die before
producing offspring. On the other hand, if such a mutation happens
in the non-coding part (where no genes are present) this does not
lead to a serious problem, and the organism produces offspring.
Therefore, such mutations are constantly accumulated in the
non-coding part and three shifted triplet distributions are mixed
into one. The fourth case (d) also produces a mix of triplet
distributions.

As a result, we have three distributions for the first case (a),
three for the second case (b) and one, symmetrical distribution
for the `non-coding' fragments (third case (c)). Because of
natural statistical deviations and other reasons, we have 7
clusters of points in the multidimensional space of triplet
frequencies.

For more illustrative material see
\cite{7clustersPPT,14Zinovyev02,14Gorban03,147clulast,147cluPhA}.

\section{Clustering and Visualizing Results}

The next step is clustering the fragments into 7 clusters. This
can be explained to the students that as a classification of
fragments of text by similarity in their triplet distributions.
This is an unsupervised classification (the cluster centers are
not known in advance).

Clustering is performed by the K-Means algorithm using the Matlab
Statistical toolbox. It is implemented in the {\it ClustFreq.m}
file. The first argument is the frequency table name and the
second argument is the number of clusters proposed. As we visually
identified 7 clusters, in this activity we put 7 as the value of
the second argument:

\vspace{15pt}

\begin{minipage}[l]{10cm}
  {\ttfamily
fragn = ClustFreq(xx3,7);}
\end{minipage}

\vspace{15pt}

\begin{figure}
\centering{
\includegraphics[width=70mm, height=70mm]{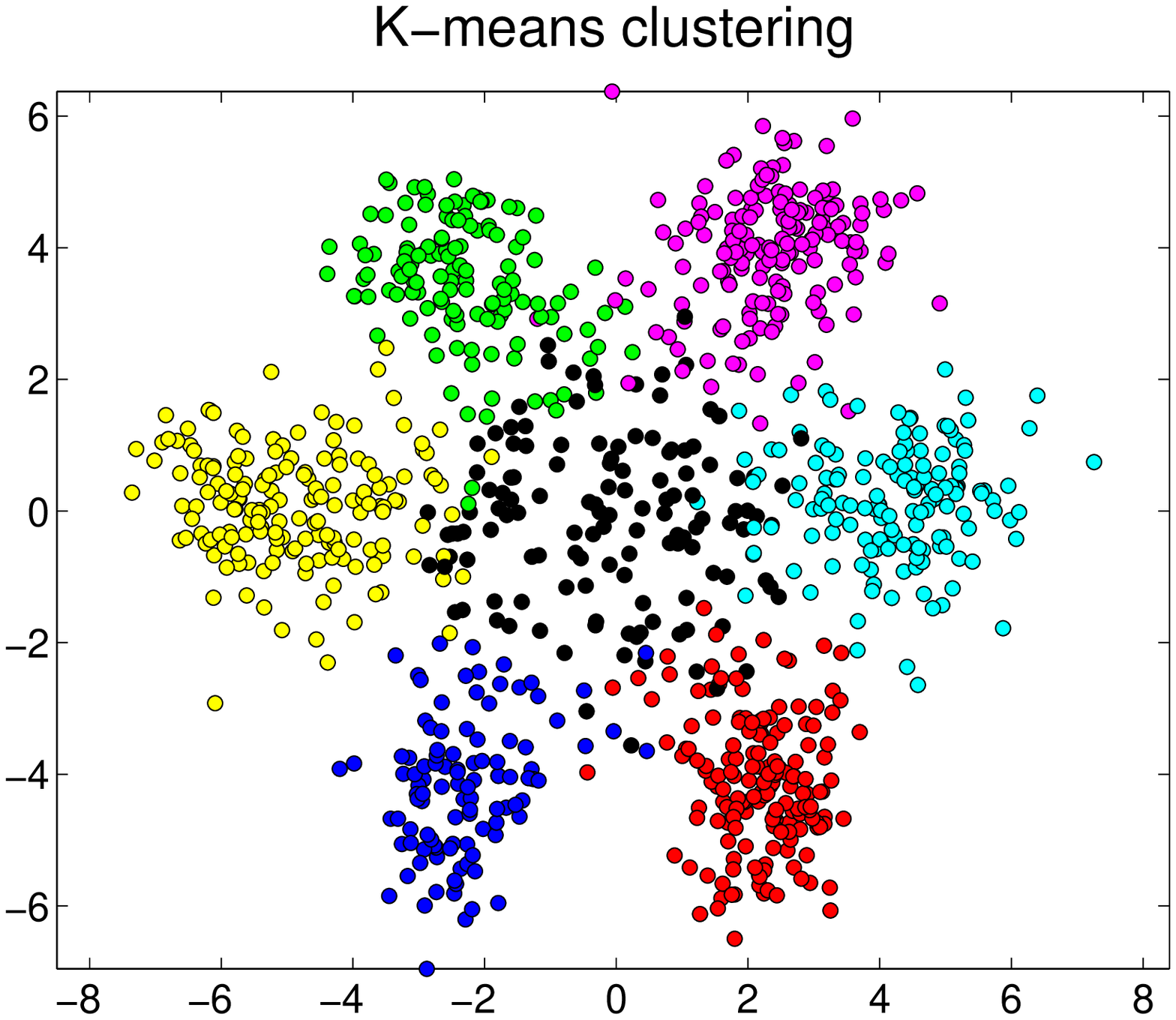}\\
\textbf{a})\\
 \includegraphics[width=115mm, height=70mm]{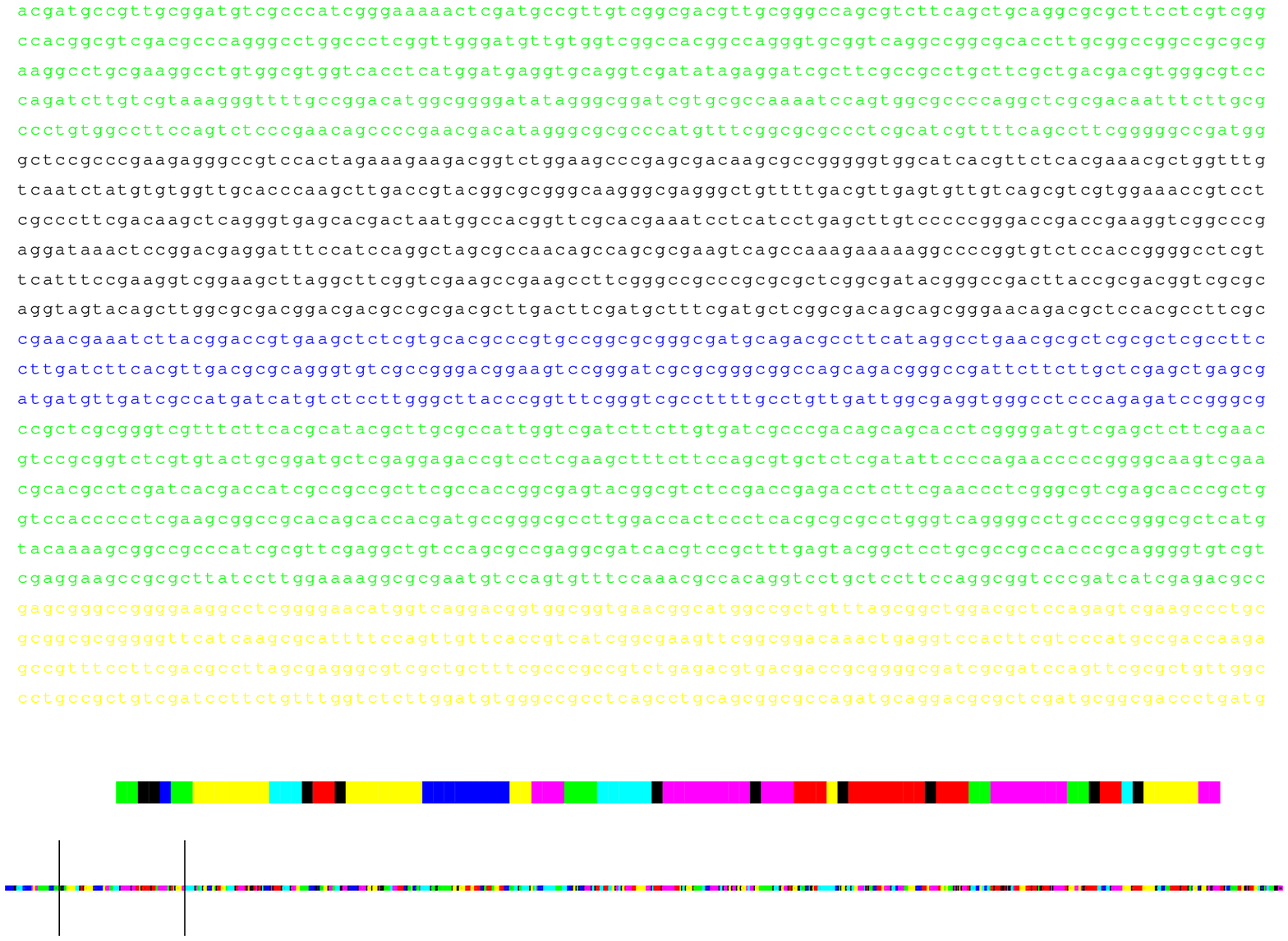}
\textbf{b})\\ }\caption{K-Means clustering of triplet frequencies
(\textbf{a}) and visualizing the clustering results onto the text
(\textbf{b}). There are three scales in the browser. The bottom
color code represents the genetic text as a whole. The second line
shows colors of 100 fragments starting from a position in the text
specified as an argument of the  {\it GenBrowser.m} function. The
letter color code shows 2400 symbols, starting from the same
position. The black color corresponds to the fragments with
non-coding information (the central cluster); other colors
correspond to locations of coding information in different
directions and with different shifts with respect to randomly
chosen division of the text into fragments \label{genbrowser}}
\end{figure}

The function assigns different colors onto the cluster points. The
cluster that is the closest to the center of the picture is
automatically colored in black (see Fig.~\ref{genbrowser}).

After clustering, every fragment of the text is assigned a cluster
label (a color). The {\it GenBrowser} function puts this color
back to the text and then visualizes it. The input arguments of
this function are a string with the genetic text, the fragment
size, a vector with cluster labels and a position in the genetic
text to read from (we use a value 13000, which can be changed for
any other position):

\vspace{15pt}
\begin{minipage}[l]{10cm}
  {\ttfamily
GenBrowser(str,300,fragn,13000);}
\end{minipage}
\vspace{15pt}

This function implements a simple {\it genome browser} (a program
for visualizing genome sequence together with other properties)
with information about gene positions.

It is explained to the students that clustering of fragments
corresponds to segmentation of the whole sequence into homogeneous
parts. The homogeneity is understood as similarity of the short
word frequency distributions for the fragments of the same cluster
(class). For example, for the following text

\vspace{10pt} {\noindent \centering{
aaaaaaaatataaaaaattttatttttttattttttggggggggggaagagggggccccccgcctccccccc}
} \vspace{10pt}

\noindent one can try to apply the frequency dictionary of length
1 for a fragment size around 5-10 to separate the text into four
homogeneous parts.

\section{Task List and Further Information}

Interested students can continue this activity. They can modify
the Matlab functions in order to solve the following problems (the
first three of them are rather difficult and require programming):

\vspace{5pt}

{\it Determine a cluster corresponding to the correct shift.} As
it was explained in the ``Understanding plots'' section, some
fragments overlap with genes in such a way that the information
can be read with the correct shift, whereas others contain the
``shifted'' information. The problem is to detect the cluster
corresponding to the correct shift. To give a hint, we can say
that the correct triplet distribution (probably) will contain the
lowest frequency of the stop codons TAA, TAG and TGA (see
\cite{Staden82-2NAR,GorbanIJCNN03}). Stop codon can appear only
once in a gene because it terminates its transcription.

\vspace{5pt} {\it Measure information content for every phase.}
The information of a triplet distribution with respect to a letter
distribution is $I = \sum_{ijk} f_{ijk}\ln
\frac{f_{ijk}}{p_{i}p_{j}p_{k}}$, where $p_i$ is a frequency of
letter $i$ (a genetic text is characterized by four such
frequencies), and $f_{ijk}$ is the frequency of triplet $ijk$.
Each cluster is a collection of fragments. Each fragment $F$ can
be divided on triplets starting from the first letter. We can
calculate the information value of this triplet distribution
$I(F)$ for each afragment $F$ . Is the information of fragments in
the cluster with a correct shift significantly different from the
information of fragments in other clusters? Is the {\it mean}
information value in the cluster with a correct shift
significantly different from the {\it mean} information value of
fragments in other clusters? Could you verify the hypothesis that
in the cluster with a correct shift the mean information value is
higher than in other clusters?

\vspace{5pt} {\it Increase resolution of determining gene
positions.} In this exercise, we use the same set of fragments to
calculate the cluster centers and to annotate the genome. As a
result the gene positions are determined with precision  that is
equal to the fragment size. In fact, cluster centers can be
calculated with non-overlapping fragments and then the genome can be
scanned again with a sliding window (this will produce a set of
overlapping fragments). Following this, triplet frequencies for each
window can be calculated and the corresponding cluster in the
64-dimensional space can be determined. The position in the text is
assigned a color corresponding to the content of the fragment
centered in this position. For further details of this process, see
\cite{GorbanGeneRecPrep01,14GorbanOpSys03,14Gorban03,Zhang03}.

\vspace{5pt} {\it Precise start and end positions of genes.} The
biological mechanism for reading genes (the polymerase molecule)
identifies the beginning and the end of a gene using special
signals, known as specialized codons. Therefore, almost all genes
start with ``ATG'' start codon and end with ``TAG'', ``TAA'' or
``TGA'' stop codons. Try to use this information to find the
beginning and end of every gene. \vspace{5pt}

{\it Play with natural texts.} The {\it CalcFreq} function is not
designed specifically for the four-letter alphabet text of genome
sequences; it can work with any text. For natural text, it is also
possible to construct local frequency dictionaries and segment it
into ``homogeneous'' (in short word frequencies) parts. But be
careful with longer frequency dictionaries, for an English text
one has $(26+1)^2=729$ possible duplets (including space as an
extra letter)! \vspace{5pt}

Students can also look at visualizations of 143 bacterial genomic
sequences at \cite{web}. All possible types of the 7-cluster
structure have been described in \cite{147clulast}. Nonlinear
principal manifolds were utilized for visualization of the 7-cluster
structure in \cite{GorbanIJCNN03}. Principal trees are applied for
visualization of the same structure in \cite{GorSumZin2007Springer}.

\section{Conclusion}

In this exercise on applying PCA and $K$-means to the analysis of a
genomic sequence, the students learn basic bioinformatics notions
and train to apply PCA to the visualization of local frequency
dictionaries in genetic text.

\newpage

\section*{Appendix 1. Program listings}

\subsection*{Function LoadSeq Listing}

\frame{

\begin{minipage}[l]{10cm}

\vspace{10pt}
 {\small  {\ttfamily
\hspace{5pt} function str=LoadSeq(fafile)

\hspace{5pt} fid = fopen(fafile); i=1; str = '';

\hspace{5pt} disp('Reading fasta-file...');

\hspace{5pt} while 1

\hspace{20pt}    if round(i/200)==i/200

\hspace{20pt}    disp(strcat(int2str(i),' lines'));

\hspace{20pt}    end

\hspace{20pt}    tline = fgetl(fid);

\hspace{20pt}    if ~ischar(tline), break, end;

\hspace{20pt}    if(size(tline)~=0)

\hspace{20pt}    if(strcmp(tline(1),'>')==0)

\hspace{20pt}    str = strcat(str,tline);

\hspace{20pt}    end;

\hspace{20pt}    end;

\hspace{20pt}    i=i+1;

\hspace{5pt} end

\hspace{5pt} nn = size(str); n = nn(2);

\hspace{5pt} disp(strcat('Length of the string: ',int2str(n))); }}

\vspace{10pt}

\end{minipage}
}

\subsection*{Function PCAFreq Listing}

\frame{

\begin{minipage}[l]{10cm}

\vspace{10pt}
 {\small  {\ttfamily
\hspace{5pt} function PCAFreq(xx)

\hspace{5pt} \% standard normalization

\hspace{5pt} nn = size(xx); n = nn(1) mn = mean(xx);

\hspace{5pt} mas = xx - repmat(mn,n,1); stdr = std(mas);

\hspace{5pt} mas = mas./repmat(stdr,n,1);

\hspace{5pt} \% creating PCA plot

\hspace{5pt} [pc,dat] = princomp(mas);

\hspace{5pt} plot(dat(:,1),dat(:,2),'k.'); hold on;

\hspace{5pt} set(gca,'FontSize',16);

\hspace{5pt} axis equal;

\hspace{5pt} title('PCA plot for text fragments','FontSize',22);

\hspace{5pt} set(gcf,'Position',[232   256   461 422]);

\hspace{5pt} hold off; }}

\vspace{10pt}
\end{minipage}
}

\newpage

\vspace{15pt}

\subsection*{Function CalcFreq Listing}

\frame{
 \begin{minipage}[l]{10cm}

\vspace{10pt}

 {\small  {\ttfamily
\hspace{5pt} function xx=CalcFreq(str,len,wid)

\hspace{5pt} disp('Cutting in fragments...');

\hspace{5pt} i=1; k=1;nn = size(str);

\hspace{5pt} while i+wid<nn(2)

\hspace{20pt}    if round(k/200)==k/200

\hspace{20pt}    disp(strcat(int2str(k),' fragments'));

\hspace{20pt}    end

\hspace{20pt}    frag = str(i:i+wid-1); vf(k) = calcf(frag,len);

\hspace{20pt}    i = i+wid; k=k+1;

\hspace{5pt} end

\hspace{5pt} disp('Merging into table...');

\hspace{5pt} names = java.util.Vector; n = 0;

\hspace{5pt} for i=1:size(vf)

\hspace{20pt}   if size(vf(i))~=0

\hspace{40pt}   keys = vf(i).keys;

\hspace{40pt}   while keys.hasMoreElements

\hspace{60pt}   key = keys.nextElement;

\hspace{60pt}  if names.indexOf(key)==-1

\hspace{60pt}  names.add(key);

\hspace{60pt}  end

\hspace{40pt}  end,  n=n+1;  end,    end

\hspace{5pt} xx = zeros(n,size(names));

\hspace{5pt} for i=1:size(vf)

\hspace{20pt}   if size(vf(i))~=0

\hspace{20pt}    if round(i/200)==i/200

\hspace{20pt}        disp(strcat(int2str(i),' points'));

\hspace{20pt}    end

\hspace{20pt}       for j=1:size(names)

\hspace{20pt}  xx(i,j) = getwf(names.elementAt(j-1),vf(i));

\hspace{20pt}       end,    end,    end

\vspace{10pt}

\hspace{5pt} function vf=calcf(str,num)

\hspace{5pt} vf = java.util.Hashtable; i = 1; nn =  size(str);

\hspace{5pt} while i+num<nn(2)

\hspace{5pt} wrd = str(i:i+num-1); i = i+num; addwf(wrd,vf,1);

\hspace{5pt} end

\vspace{10pt}

\hspace{5pt} function addwf(word,hash,fr)

\hspace{5pt} wf = hash.get(word);

\hspace{5pt} if size(wf)==0 hash.put(word,fr); else

\hspace{5pt} hash.put(word,fr+wf); end

\vspace{10pt}

\hspace{5pt} function fr=getwf(word,hash)

\hspace{5pt} wf = hash.get(word);

\hspace{5pt} if size(wf)==0 r=0; else fr=wf; end

} }

\vspace{10pt}

\end{minipage} }

\newpage

\subsection*{Function ClustFreq Listing}

\frame{

\begin{minipage}[l]{10cm}

\vspace{10pt}

{\small  {\ttfamily

\hspace{5pt} function fragn = ClustFreq(xx,k)

\hspace{5pt} \% centralization and normalization

\hspace{5pt} nn = size(xx); n = nn(1)

\hspace{5pt} mn = mean(xx);

\hspace{5pt} mas = xx - repmat(mn,n,1);

\hspace{5pt} stdr = std(mas);

\hspace{5pt} mas = mas./repmat(stdr,n,1);

\hspace{5pt} \% calculating principal components

\hspace{5pt} [pc,dat] = princomp(mas);

\hspace{5pt} \% k-means clustering

\hspace{5pt} [fragn,C] = kmeans(mas,k);

\hspace{5pt} \% projecting cluster centers into the PCA basis

\hspace{5pt} XTP = C; temp = size(XTP); nums = temp(1);

\hspace{5pt} X1c = XTP-repmat(mn,nums,1);

\hspace{5pt} X1r = X1c./repmat(stdr,nums,1);

\hspace{5pt} X1P = pc'*X1r'; X1P = X1P';

\hspace{5pt} \% marking the central cluster black

\hspace{5pt} cnames = ['k','r','g','b','m','c','y'];

\hspace{5pt} for i=1:k  no(i) = norm(X1P(i,1:3)); end

\hspace{5pt} [m,mi] = min(no);

\hspace{5pt} for i=1:size(fragn)

\hspace{20pt}      if fragn(i)==mi fragn(i)=1;

\hspace{20pt}      elseif fragn(i)==1 fragn(i)=mi; end

\hspace{5pt} end

\hspace{5pt} \% plotting the result using PCA

\hspace{5pt} for i=1:n

\hspace{20pt}    plot(dat(i,1),dat(i,2),'ko',

\hspace{20pt} 'MarkerEdgeColor',[0 0 0],'MarkerFaceColor',

\hspace{20pt} cnames(fragn(i)));

\hspace{20pt} hold on;

\hspace{5pt} end

\hspace{5pt} set(gca,'FontSize',16); axis equal;

\hspace{5pt} title('K-means clustering','FontSize',22);

\hspace{5pt} set(gcf,'Position',[232 256   461 422]);

}}

\vspace{10pt}

\end{minipage}
}
\newpage

\vspace{15pt}

\subsection*{Function GenBrowser Listing}

\frame{

\begin{minipage}[l]{10cm}

\vspace{10pt}

{\small  {\ttfamily

\hspace{5pt} function GenBrowser(str,wid,fragn,startp)

\hspace{5pt} \% we will show 100 fragments in the detailed view

\hspace{5pt} endp = startp+wid*100; nn = size(fragn); n = nn(1);

\hspace{5pt} xr1 = startp/(n*wid); xr2 = endp/(n*wid);

\hspace{5pt} cnames = ['k','r','g','b','m','c','y'];

\hspace{5pt} subplot('Position',[0 0 1 0.1]);

\hspace{5pt} for i=1:size(fragn)

\hspace{15pt}
plot(i/n,0,strcat(cnames(fragn(i)),'s'),'MarkerSize',2);

\hspace{15pt} hold on;

\hspace{5pt} end

\hspace{5pt} plot([xr1 xr1],[-1 1],'k'); hold on;

\hspace{5pt} plot([xr2 xr2],[-1 1],'k'); axis off;

\hspace{5pt} subplot('Position',[0 0.1 1 0.1]);

\hspace{5pt} for i=floor(startp/wid)+1:floor(endp/wid)+1

\hspace{15pt}    plot([(i-0.5)*wid (i+0.5)*wid],[0 0],

\hspace{15pt}    strcat(cnames(fragn(i)),'-'),'LineWidth',5);

\hspace{15pt} hold on;

\hspace{5pt} end

\hspace{5pt} axis off;

\hspace{5pt} subplot('Position',[0 0.25 0.98 0.75]);

\hspace{5pt} xlim([0,1]); ylim([0,1]); twid = 100; nlin = 24;k=startp;

\hspace{5pt} for j=1:nlin

\hspace{5pt} for i=1:twid

\hspace{15pt}    col = cnames(fragn(floor(k/wid)+1));

\hspace{15pt} h=text(i/twid,1-j/nlin,str(k),'FontSize',8,

\hspace{15pt} 'FontName','FixedWidth');

\hspace{15pt}    set(h,'Color',col);

\hspace{15pt}    k=k+1;

\hspace{5pt} end end

\hspace{5pt} axis off;

\hspace{5pt} set(gcf,'Position',[64   356   879   195]);

}}

\vspace{10pt}

\end{minipage}
}
\newpage

\section*{Appendix 2. PCA plots of triplet frequencies}

\begin{figure}
a)\includegraphics[width=33mm]{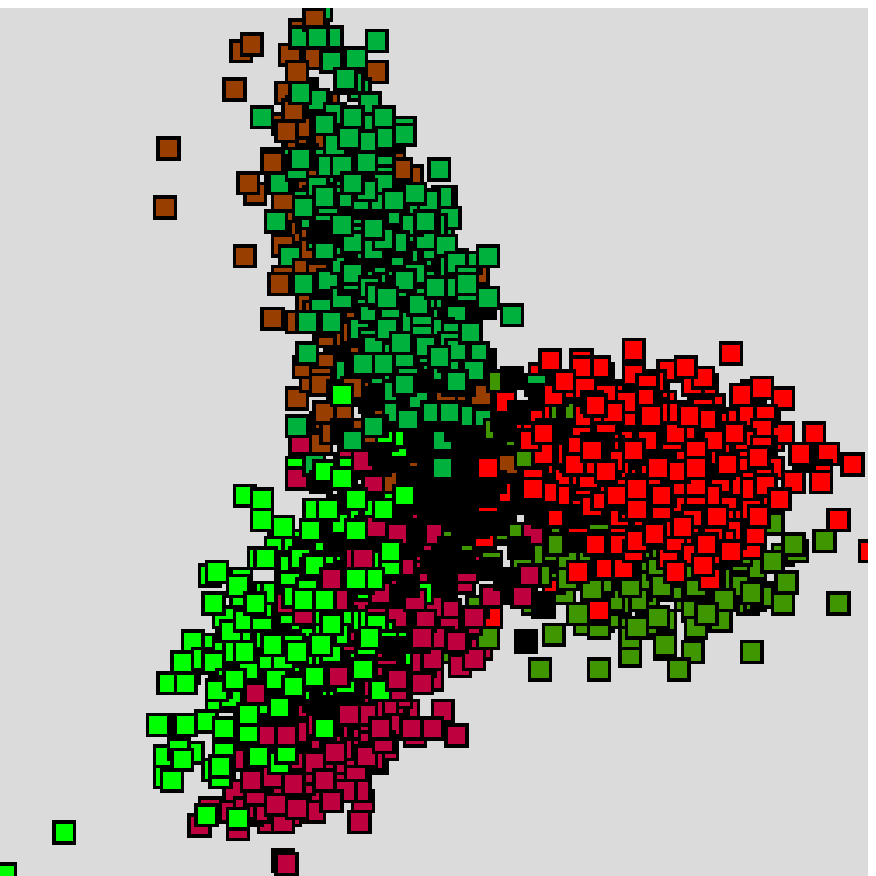}
b)\includegraphics[width=33mm]{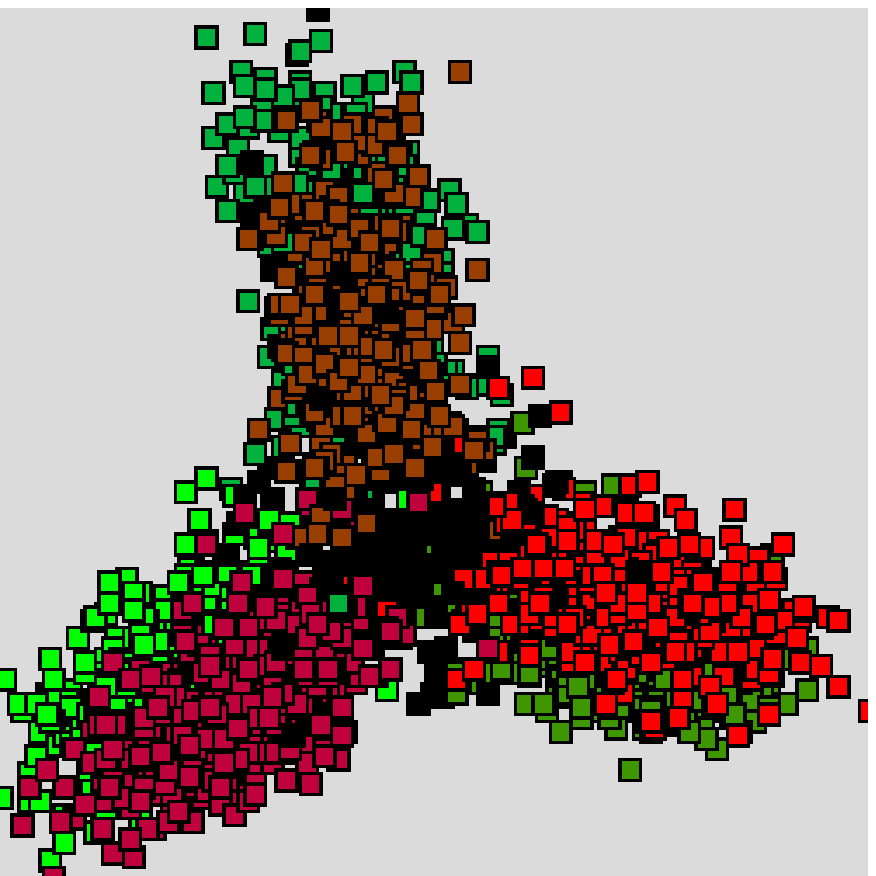}
c)\includegraphics[width=33mm]{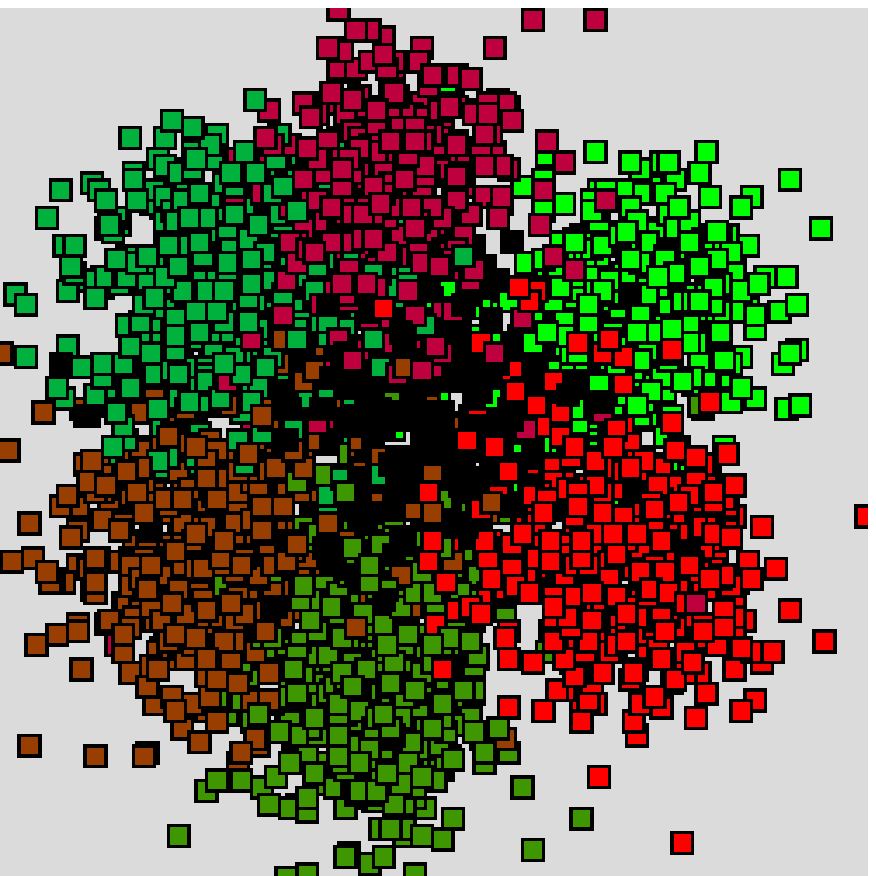}
\includegraphics[width=6mm]{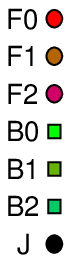}\\
d)\includegraphics[width=33mm]{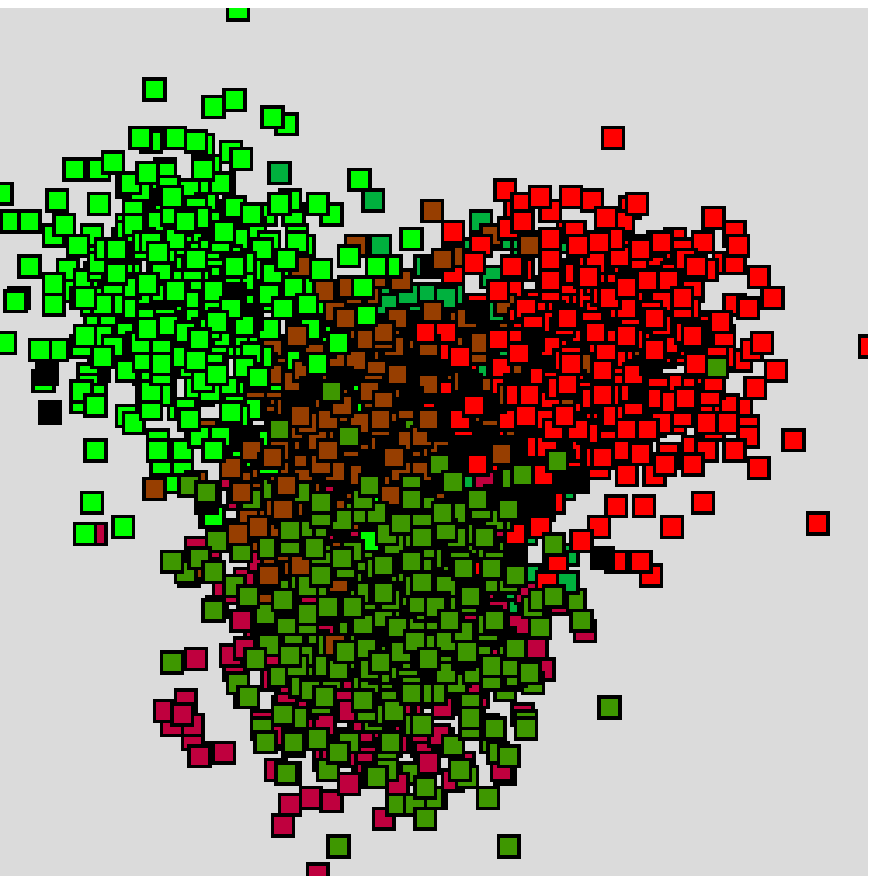}\
e)\includegraphics[width=33mm]{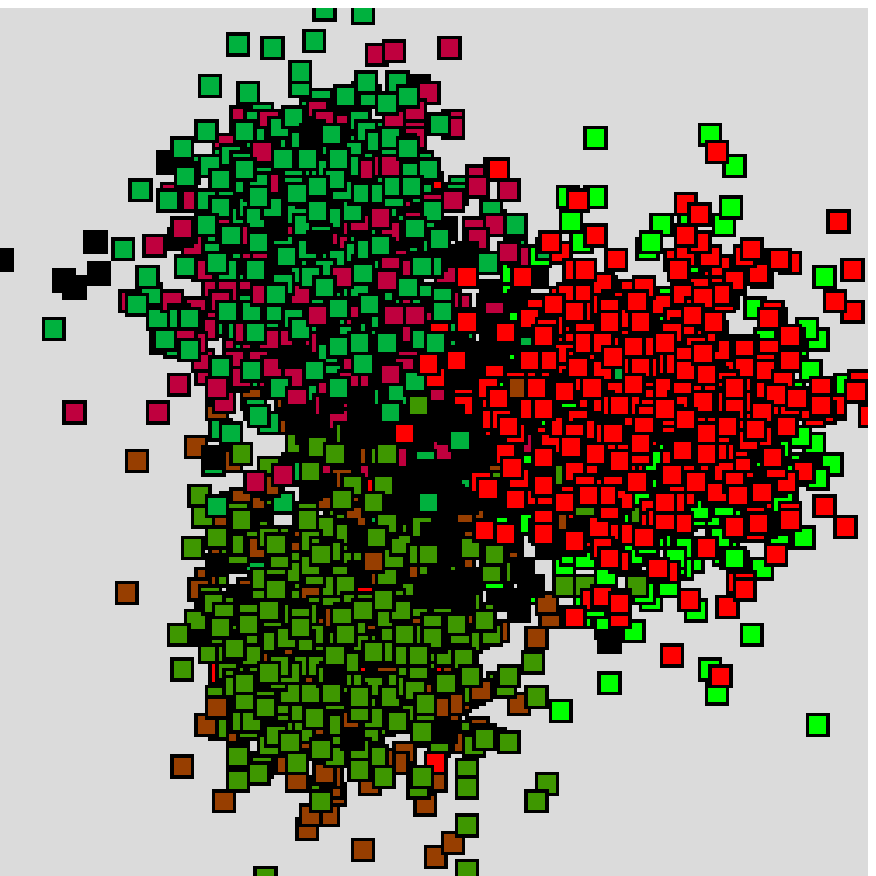}\
f)\includegraphics[width=33mm]{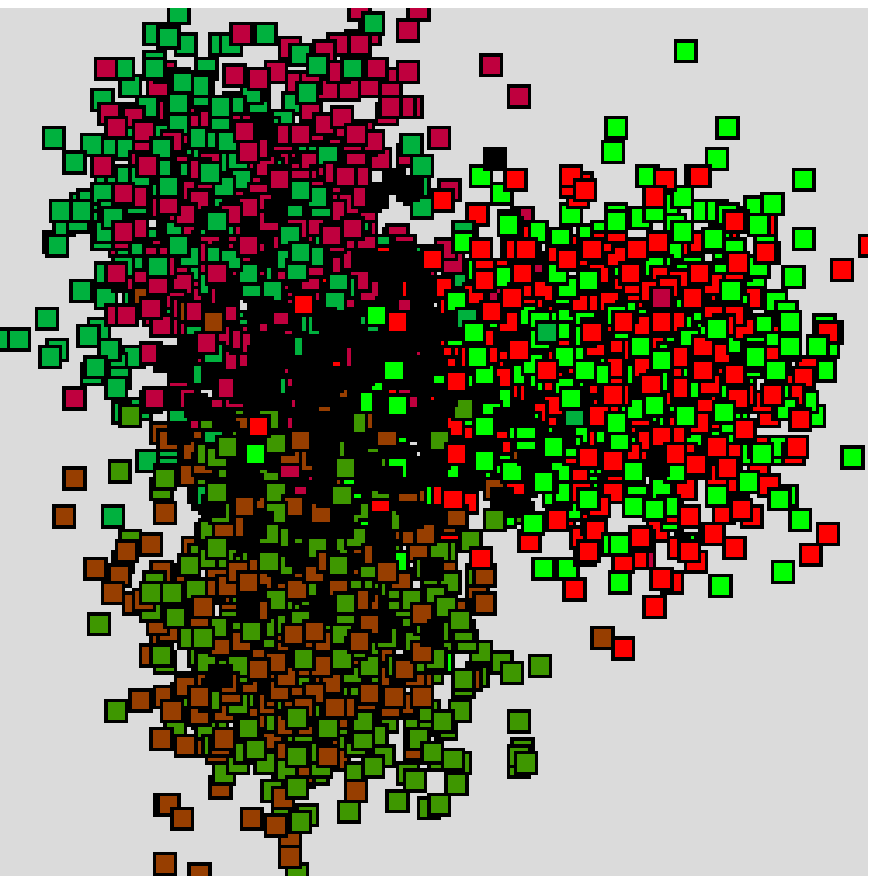}\\
g)\includegraphics[width=33mm]{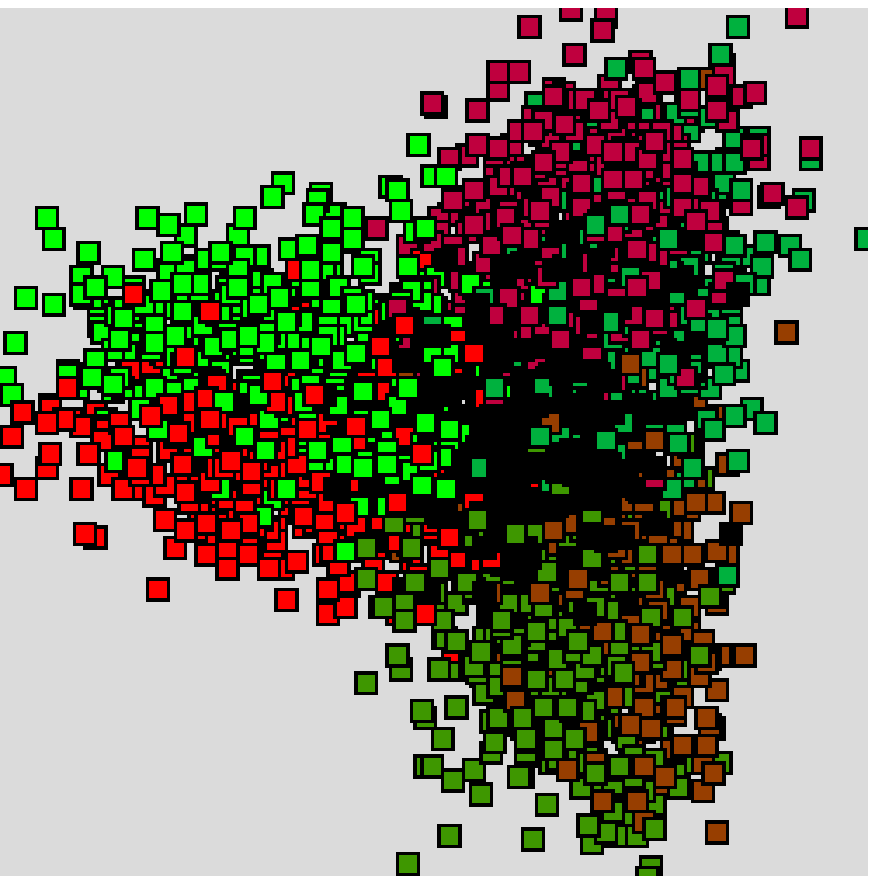}\
h)\includegraphics[width=33mm]{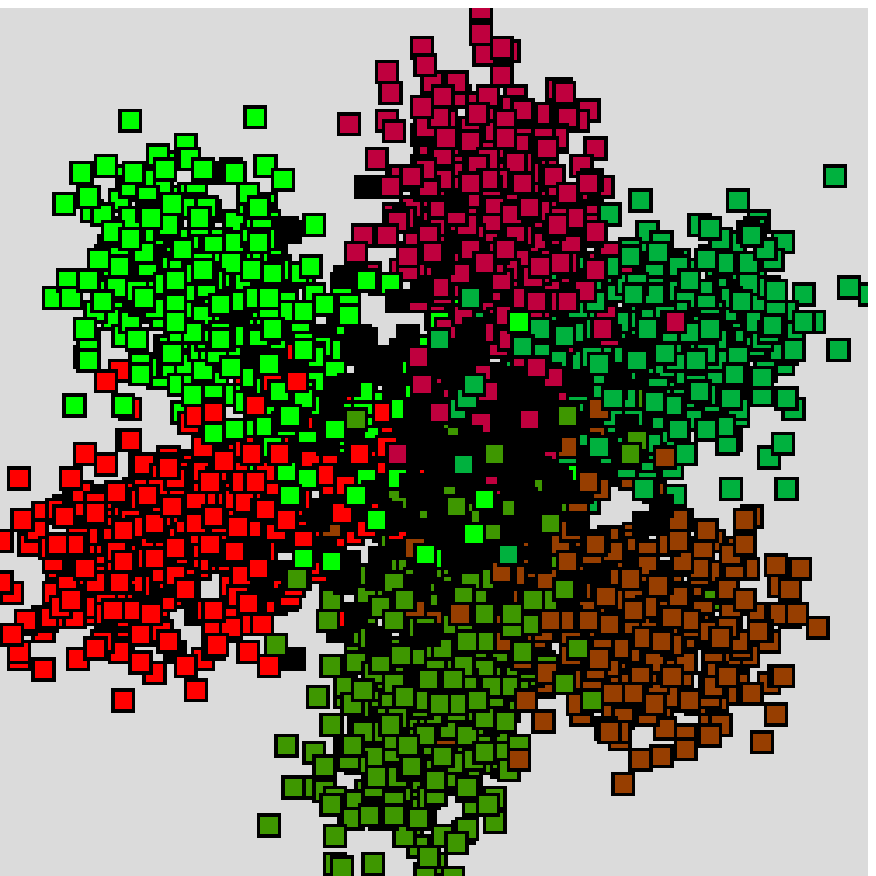}\
i)\includegraphics[width=33mm]{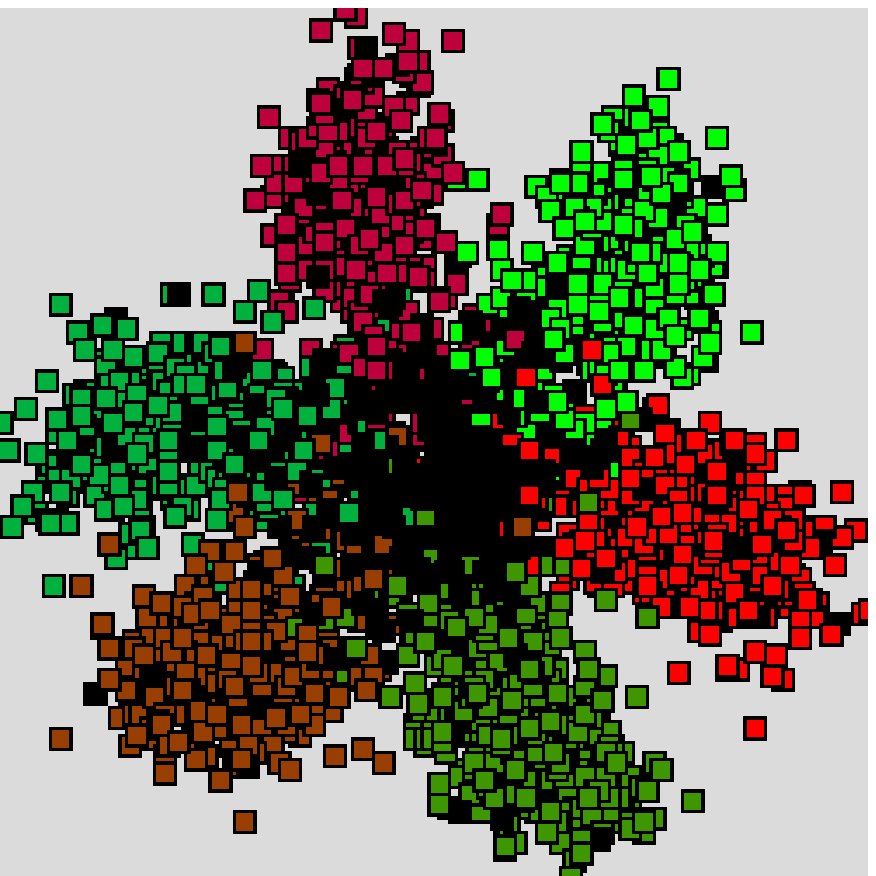}
\caption{PCA plots of triplet frequencies. Typical 7-cluster
structures: (a) Ureaplasma urealyticum, Bacteria Firmicutes
Bacillus/Clostridium group; Code: Uu,   GC-content: 0.25, Length:
751719 bp; (b) Fusobacterium nucleatum, Bacteria,   Fusobacteria,
Fusobacterium; Code: Fn, GC-content: 0.27,   Length: 2174500 bp;
(c) Helicobacter pylori 26695, Bacteria, Proteobacteria, epsilon
subdivision; Code: Hp, GC-content: 0.39, Length: 1667867 bp; (d)
Bacillus halodurans, Bacteria, Firmicutes Bacillus/Clostridium
group; Code: Bh, GC-content: 0.44, Length: 4202353 bp; (e) Vibrio
cholerae, Bacteria, Proteobacteria, gamma subdivision; Code: Vc,
GC-content: 0.48, Length: 2961149 bp; (f) Escherichia coli K12;
Bacteria, Proteobacteria, gamma subdivision; Code: Ec, GC-content:
0.51,   Length: 4639221 bp; (g) Corynebacterium glutamicum,
Bacteria, Firmicutes, Actinobacteria; Code: Cg, GC-content: 0.54,
Length: 3309401 bp; (h) Bifidobacterium longum, Bacteria,
Actinobacteria, Actinobacteridae; Code: Bl, GC-content: 0.6,
Length: 2256646 bp; (i) Streptomyces coelicolor, Bacteria,
Firmicutes, Actinobacteria; Code: Sc, GC-content: 0.72, Length:
8667507 bp. \label{PCAplots}}
\end{figure}
The observed cluster structure clearly depends on the GC-content.
Genomes in Fig.~\ref{PCAplots} are ordered from GC-poor genomes
(GC-content 0.25 for Ureaplasma urealyticum) to GC-rich ones
(GC-content 0.72 for Streptomyces coelicolor).

In short, this is a PCA plot of the point distribution. Every
point on this plot presents a fragment of the genetic text,
characterized by 64-dimensional vector. The length of these
fragments is 300 bp. Each fragment is cut into 100 non-overlapped
triplets (starting from the first position). The 64 components of
the vector are frequencies of different triplets. These points are
colored. The colour shows the biological meaning of the fragment.
The DNA double helix consists of two strands: the leading strand
and the lagging strand. In Fig.~\ref{PCAplots} F0 stands for the
the group ''coding'' fragments of F-type (genes in leading strand,
``F" stands for ``Forward") in which nonoverlapping triplets have
been red in the correct frame. F1 and F2 correspond to the
fragments where the triplets have been red with a frameshift (on
one or two positions). Analogously, the B0, B1 and B2 labels stand
for the B-type fragment groups (genes in lagging strand, ``B"
stands for ``Back"), where the triplets have been red with one of
three possible frameshifts (0, 1, 2), respectively. Principal
component analysis allows to represent the 64-dimensional point
distribution on 2D-plane and, thus, visualize its cluster
structure.

Let us describe the basic properties of the structure. First, it
consists of seven clusters.This fact is rather natural. Indeed, we
clip fragments only from the forward strand and every fragment can
contain (1) piece of coding region from the forward strand, with
three possible shifts relatively to the first fragment position;
(2) coding information from the backward strand, with three
possible frameshifts; (3) non-coding region; (4) mix of coding and
non-coding information: these fragments introduce noise in our
distribution, but their relative concentration is not high.

Second, the structure is rather well pronounced. This means that
most of learning (and even self-learning) techniques aiming at
separation of the clusters from each other will work very well,
which is the case for bacterial gene-finders that have performance
more than 90\% in most cases (for recent overview, see
\cite{Mathe}).

Third, the structure is well represented by a 3D-plot (and even
presented 2D plots give the proper picture). These animated 3D
plots for all genomes from Fig.~\ref{PCAplots} are added to the
preprint as separate gif-files.

Forth, it is indeed has symmetric and appealing patterns, that
clearly depend on GC-content. This structure is explained in our
paper \cite{147cluPhA}, some additional details for more genomes
are presented in \cite{mystery}, and PCA analysis of many genomes
is presented online on the special web-site \cite{web}.

\end{document}